\documentclass[12pt,a4]{article}

\usepackage{amssymb}
\usepackage{amsmath}
\usepackage{graphicx}

\begin{document}

\title{Aharonov-Bohm rings in de Sitter expanding universe}

\author{Ion I. Cot\u aescu\\
{\small \it West University of Timi\c{s}oara,}\\
{\small \it V.  P\^{a}rvan Ave.  4, RO-300223 Timi\c{s}oara, Romania}}

\maketitle

\begin{abstract}
The $(1+3)$-dimensional Dirac equation of the fermions moving in ideal Aharonov-Bohm rings in the de Sitter expanding universe is used for deriving the exact expressions of the general relativistic partial currents and corresponding energies. In the de Sitter geometry, these quantities depend on time but these are related each other just as  in the non-relativistic case  or in special relativity.  A specific relativistic effect is the saturation of the partial currents for high values of the total angular momentum. The total relativistic persistent current at $T=0$ takes over this property even though it is evolving in time because of the de Sitter expansion.   
 
\end{abstract}


\newpage

\section{Introduction}

The electronic effects in mesoscopic systems can be studied  in many cases by restricting to the non-relativistic quantum mechanics \cite{B1,B2,B3,B4,B5,B6,B7,B8,B9,B10}  based on the Schr\" odinger  equation with suitable  additional terms describing the spin-orbit interaction \cite{B11,B12,B13,B14,B15,B16}. Nevertheless, there are nano-systems laying out some measurable relativistic effects that can be satisfactory explained  considering the electrons as  massless Dirac particles moving on lattices \cite{N1,N2,Y1,Y2}.  In many  studies concerning relativistic effects in  nano-systems \cite{N1,N2, B17,B18,B18a,B20} one considers the electrons near the Fermi surface as  $(1+2)$-dimensional Dirac particles corresponding to a restricted three-dimensional Clifford algebra. However, in this manner one restricts simultaneously not only the orbital degrees of freedom but the spin ones too, reducing the natural $SO(1,4)$ symmetry to the $SO(1,2)$ one.   

We believe that this restriction is forced and less justified especially when one can use the {\em complete} $(1+3)$-dimensional Dirac equation  restricting only the orbital motion, according to the concrete geometry of the studied system, but without affecting the natural spin degrees of freedom described by the $SL(2,{\Bbb C})$ group. Nevertheless, the complete  Dirac equation was only occasionally used for investigating  some special problems of the fermions in external Aharonov-Bohm (AB) field as for example the spin effects in perturbation theory \cite{scat1,scat2,scat3}, the behaviour of the AB  fermions  in MIT cylinders \cite{MIT} and even the AB dynamics using numerical methods.  

Trying to apply the complete Dirac equation for studying the relativistic persistent currents in AB quantum rings we obtained initially a version of Dirac equation  with a non-Hermitian term  introduced by the unappropriate method of orbital restriction \cite{CP}. This  distorts the results presented  therein  as well as other ones based on this approach,  even though these may outline some new realistic effects  \cite{Ghosh}. For this reason, we corrected this inconvenience building the complete theory of the relativistic AB effect on ideal rings \cite{CAB1} and cylinders \cite{CAB2} considering  the {\em correctly restricted} $(1+3)$-dimensional Dirac equation in the Minkowski spacetime, involving only Hermitian operators  that can be obtained easily starting with a suitable restricted Lagrangian theory \cite{CAB1}.

In this paper we would like to extend this study to the expanding portion of the de Sitter (dS) spacetime which is a plausible model of our universe known as the dS expanding universe. On this manifold  we studied the solutions of the free Dirac equation \cite{CdS1} and we constructed the quantum electrodynamic in Coulomb gauge analyzing simple effects in the first order of the perturbation theory  \cite{CQED}. Here we focus on  the relativistic AB effect in ideal rings in the presence of the dS gravity showing how the orbital motion can be restricted without affecting the spin degrees of freedom. We derive the fundamental solutions of the Dirac equations as common eigenspinors of the set of commuting operators formed by the Dirac operator, the total angular momentum along the direction of the magnetic field and a special operator analogous to the Dirac spherical operator of the relativistic central problems \cite{TH}. These solutions can be normalized with respect to the relativistic scalar product obtaining  the system of normalized fundamental solutions that allow us to write down the exact expressions of the energy, relativistic partial currents and to derive the persistent persistent current at $T=0(K)$.   

The principal feature of this problem is that the energy and the partial current densities are no longer conserved depending on time by means of the time-modulation functions of the fundamental spinors. This fact is not surprising but is quite new such that we derive the closed expressions of the energy spectrum and the form of the partial currents. The surprise is to find that these quantities are related just as in Minkowski spacetime \cite{CAB1} or even in the non-relativistic case. More specific, we deduce that the partial currents are proportional with the derivative of the energy with respect to the flux parameter. Moreover, we point out the same saturation effect we discovered in special relativity but which is missing in the non- relativistic limit \cite{CAB1}. 

In our approach the time-dependence of the principal quantities is given by complicated functions that may lead to difficulties in  interpreting the final results. For this reason, we propose a satisfactory approximation which helps us to write down simpler and intuitive formulas we need for understanding the phenomenology as well as the limiting cases. In this manner we can show that in the flat limit the dS persistent currents become just those found in the Minkowski flat spacetime \cite{CAB1}.  

The paper is organized as follows. In the second section we present the relativistic theory of the fermions in AB rings based on a suitable restricted complete Dirac equation deducing the form of the normalized spinors of given total angular momentum and polarization. In the next section we deduce the expression of the time-dependent energy expectation values related to the partial currents which also are depending on time.  After we verify the mentioned relation the energy and partial current, we study the functions determining their time evolution. In section 4 we propose an approximation of these functions showing by brief graphical analyze that this is accurate inducing relative errors less than $10^{-6}$. In this approximation the physical quantities, i. e. the persistent current and the energy of the Fermi level,  are given by intuitive formulas allowing us to discuss their physical  meaning in the last section. 

We work in natural units with $\hbar=c=1$ resorting to the SI ones in some concrete examples.

\section{AB rings in dS spacetime}

The de Sitter (dS) expanding universe is the expanding portion of the $(1+3)$-dimensional dS hyperboloid where we can  use the well-known types of co-moving charts, either  the conformal chart $\{t_c, \vec{x}\}$ whose coordinates $x^{\mu}$ ($\alpha,..\mu,\nu...=0,1,2,3$) are the conformal time $x^0=t_c\in(-\infty, 0]$ and the Cartesian space coordinates $x^i$ ($i,j,k...=1,2,3)$, or the FLRW chart $\{t, \vec{x}\}$  with the proper (or cosmic) time $t\in (-\infty,\infty)$, related to the conformal one as 
\begin{equation}
\omega t_c =-e^{-\omega t}\,,
\end{equation}
where $\omega$ is the Hubble dS constant in our notation. The line elements of these charts   \cite{BD},
\begin{equation}\label{line}
ds^{2}=\frac{1}{\omega^2 {t_c}^2}\left({dt_c}^{2}-d\vec{x}\cdot d\vec{x}\right)=dt^2-e^{2\omega t}(d\vec{x}\cdot d\vec{x})\,,
\end{equation}
allow the simple choice of the diagonal tetrad gauge fields whose  non-vanishing components,   
\begin{equation}\label{tt}
e^{0}_{0}=-\omega t_{c}\,, \quad e^{i}_{j}=-\delta^{i}_{j}\,\omega t_c
\,,\quad
\hat e^{0}_{0}=-\frac{1}{\omega t_{c}}\,, \quad \hat e^{i}_{j}=-\delta^{i}_{j}\,
\frac{1}{\omega t_c}\,.
\end{equation}
give the  Dirac operator 
\begin{eqnarray}\label{ED1}
{\cal E}_D&=&-i\omega t_c\left(\gamma^0\partial_{t_{c}}+\gamma^i\partial_i\right)
+\frac{3i\omega}{2}\gamma^{0} \nonumber\\
&=&i\gamma^0\partial_{t}+ie^{-\omega t}\gamma^i\partial_i
+\frac{3i\omega}{2}\gamma^{0}\,.
\end{eqnarray}
In general, the solutions of the Dirac equation, $({\cal E}_D-m)\psi=0$, are linear combinations of  the fundamental solutions which have to form an orthonormal basis with respect to  relativistic scalar product  \cite{BD},
\begin{equation}\label{sp}
\left< \psi,\psi^{\prime}\right>=\int d^{3}x\, \mu(x)
\overline{\psi}(x)\gamma^{0}\psi^{\prime}(x) \,, \quad \overline{\psi}=\psi^+\gamma^0 \,.
\end{equation}
defined by the weight function, 
\begin{equation}\label{mu}
\mu=\sqrt{g}\,e_{0}^{0}=(-\omega t_{c})^{-3}=e^{3\omega t}\,, \quad g=|{\rm det}(g)|\,.
\end{equation}

Now we consider the ideal Aharonov-Bohm effect assuming that the Dirac fermions of mass $m$  move in an {\em ideal} ring of radius $R$  whose axis is oriented along the homogeneous and static external magnetic field $\vec{B}$. We remind the reader that the Maxwell equations are conformally invariant such that the electromagnetic potentials in the conformal chart $\{t_c,\vec{x}\}$ are just the Minkowskian ones, $A_0=0$ and $\vec{A}=\frac{1}{2}\vec{B}\land \vec{x}$.  Moreover,  the transformation $t_c\to t$ does not affect the space components of the  tensor strength such that the potentials giving the magnetic field remain the same.  

The ideal ring is a one-dimensional manifold (without internal structure) embedded in the three-dimensional space according to the equations  $r=R$ and $z=0$,  written in cylindrical coordinates $(t,\vec{x}) \to (t, r, \phi, z)$ with the $z$ axis oriented along $\vec{B}$. Then, it is natural to assume that any field $\psi$ defined on this manifold depends only on the remaining coordinates $(t,\phi)$ such that we may take $\partial_r\psi=0$ and $\partial_z\,\psi=0$ but in the Lagrangian instead of the field equation for preserving  the Hermitian properties of the Dirac operator \cite{CAB1}.  We have shown that this procedure reduces to the substitution 
\begin{equation}
\gamma^i \partial_i \to \gamma^{\phi}(\partial_{\phi}+i\beta)+\frac{1}{2}\,\partial_{\phi}(\gamma^{\phi})\,,
\end{equation}
where 
\begin{equation}\label{gamph}
\gamma^{\phi}=\frac{1}{R}(-\gamma^1\sin\phi+\gamma^2\cos\phi)
\end{equation}
is depending on $\phi$ while the notation $\beta=\frac{1}{2}eBR^2$ stands for the usual dimensionless flux parameter (in natural units). In this manner we obtain the Dirac operator of the AB effect in the dS expanding universe, 
\begin{equation}\label{D2}
E_D=-i\omega t_c\left[\gamma^0\partial_{t_c} +\gamma^{\phi}(\partial_{\phi}+i \beta)+\frac{1}{2}\,\partial_{\phi}(\gamma^{\phi})\right] +\frac{3i\omega}{2}\gamma^{0}\,,
\end{equation}
whose supplemental third term guarantees the desired Hermitian property,  $\overline{E}_D=E_D$. Note that in the dS geometry the AB interaction term $\omega \beta \gamma^{\phi} t_c$ depends explicitly on time in contrast to the flat case where this is time-independent \cite{CAB1}.

The next step is to  look for a system of commuting operators able to determine the fundamental solutions of the Dirac equation $(E_{D}-m)\psi=0$. As in the flat case, the Dirac operator  commutes with the third component, $J_3=L_3+S_3$, of the total angular momentum, formed by the orbital part $L_3=-i\partial_{\phi}$ and the spin term $S_3=\frac{1}{2}\,{\rm diag} (\sigma_3,\sigma_3)$. Another operator commuting with $E_D$ is the angular Dirac operator that in the two-dimensional case has the simple form $K=2\gamma^0 S_3$ \cite{Dong,CAB1}. We collected thus the system of commuting operators $\{E_D,J_3,K\}$ which determines the system of fundamental solutions $U_{\lambda}^{\pm}(x)$ as common eigenspinors satisfying the eigenvalues problems $E_DU_{\lambda}^{\pm}=mU_{\lambda}^{\pm}$, $J_3U_{\lambda}^{\pm}=\lambda U_{\lambda}^{\pm}$ and $K U_{\lambda}^{\pm}=\pm U_{\lambda}^{\pm}$.

Solving these equations as in Ref. \cite{CAB1} we obtain the fundamental spinors
\begin{equation}
U_{\lambda}^+(t_c, \phi)=N_{\lambda}(\omega t_c)^2\left(
\begin{array}{c}
K_{\nu_-}(-i \zeta)    e^{i\phi(\lambda-\frac{1}{2})}\\
0\\
0\\
i K_{\nu_+}(-i\zeta)e^{i\phi(\lambda+\frac{1}{2})}
\end{array}\right) \,,
\end{equation}
\begin{equation}
U_{\lambda}^-(t_c, \phi)=N_{\lambda}(\omega t_c)^2\left(
\begin{array}{c}
0\\
K_{\nu_-}(-i \zeta)    e^{i\phi(\lambda+\frac{1}{2})}\\
-i K_{\nu_+}(-i\zeta)e^{i\phi(\lambda-\frac{1}{2})}\\
0
\end{array}\right) \,,
\end{equation}
in terms of modified Bessel functions, $K_{\nu_{\pm}}$,  of indices $\nu_{\pm}=\frac{1}{2}\pm i\frac{m}{\omega}$, depending on time through the auxiliary variable
\begin{equation}\label{zeta}
\zeta=-\frac{\lambda+\beta}{R}\, t_c=\frac{\lambda+\beta}{R\omega}\, e^{-\omega t}\,.
\end{equation}
It remains to determine the normalization constant $N$ assuming that the fundamental spinors obey the ortonormaization rules 
\begin{equation}
\langle U^{\pm}_{\lambda}, U^{\pm}_{\lambda'}\rangle=\delta_{\lambda,\lambda'}\,,\quad \langle U^{\pm}_{\lambda}, U^{\mp}_{\lambda'}\rangle=0\,,
\end{equation}
with respect to the relativistic scalar product of the AB problem
\begin{equation}\label{spAB}
\langle \psi, \psi'\rangle=R(-\omega t_c)^{-3}\,\int_{0}^{2\pi}d\phi\,\psi^{\dagger}(t_c,\phi) \psi'(t_c,\phi)\,,
\end{equation} 
resulted from the general form (\ref{sp}) after the dimensional reduction. This can be calculated according to Eq. (\ref{H3}) finding  the common normalization constant
\begin{equation}
N_{\lambda}=\frac{1}{ \pi R}\sqrt{\frac{\lambda +\beta}{2\omega}}\,.
\end{equation}

The principal quantum number is the angular one,  $\lambda$, which can take only half-integer values, $\lambda=\pm\frac{1}{2}\pm \frac{3}{2},...$,  as it results from the condition $U_{\lambda}^{\pm}(t, \phi+2\pi)=U_{\lambda}^{\pm}(t, \phi)$. The eigenvalues $\pm 1$ of the operator $K$ give the polarization in the non-relativistic limit. For this reason we keep this terminology considering that these eigenvalues  define the fermion polarization with respect to the direction of the magnetic field $\vec{B}$.  In general, a state of total angular momentum  $\lambda$ and arbitrary polarization is given by the normalized linear combination 
\begin{equation}\label{psil}
\psi_{\lambda}=c_+U^+_{\lambda}+c_-U^-_{\lambda}\,, \quad |c_+|^2+|c_-|^2=1\,,
\end{equation} 
for which  the expectation value of the polarization operator,
\begin{equation}
\langle \psi_{\lambda}, K\psi_{\lambda}\rangle =|c_+|^2-|c_-|^2\,,
\end{equation} 
can take values in the domain $[-1,1]$.

\section{Energy and  currents}
 
In the AB ring in the dS spacetime each electron produces a partial current which will contribute to the final form of the persistent current at $T=0$. In the non-relativistic approach as well as in the relativistic theory of AB rings  in Minkowski spacetime \cite{CAB1,CAB2} the partial currents are proportional with the derivative of energy with respect to $\beta$. Therefore, the first step is to investigate if this fundamental relation  holds in the dS geometry too. 

The energy operator of the dS isometry group, $H=-i \omega(t_c \partial_{t_c}+x^i\partial_i)$ \cite{CdS1}, reduces to its first term on AB rings since the dimensional reduction imposes $x^i\partial_i=r\partial_r+z\partial_z=0$. The principal difficulty arising here  is that this operator does not commute with $E_D$ such that the energy is no longer a conserved quantity as in the flat case.  Therefore we must consider the time-dependent expectation values of the operator $H_{AB}=-i \omega t_c \partial_{t_c}$ in the states $(\lambda, \pm 1)$, defined as
\begin{equation}
E_{\lambda}^{\pm}(t)= \langle U_{\lambda}^{\pm},H_{AB}U_{\lambda}^{\pm}\rangle\,.
\end{equation}
Then, by using the scalar product (\ref{spAB}),  we find that the energy is independent on polarization, depending on time and $\lambda$ only as function of $\zeta$,
\begin{equation}
E_{\lambda}^+(t_c)=E_{\lambda}^-(t_c)=E_{\lambda}(t_c)\equiv\Re E[\zeta(t_c,\lambda)]\,,
\end{equation}
where the function
\begin{eqnarray}\label{El}
E(\zeta)&=&-\frac{i \omega \zeta}{\pi}\left\{K_{\nu_+}(i\zeta)\left[ i \zeta K_{\nu_+}(-i \zeta)+ (1+\nu_{+})K_{\nu_-}(-i \zeta) \right]\right.\nonumber\\
&&~~~~~~\left.+\,K_{\nu_-}(i\zeta)\left[ i \zeta K_{\nu_-}(-i \zeta)+ (1+\nu_{-})K_{\nu_+}(-i \zeta) \right]\right\}\,.
\end{eqnarray}
has an imaginary part,  
\begin{equation}\label{ImE}
\Im E(\zeta)=-\frac{3\omega}{2}\,{\rm sign}( \zeta)\,.
\end{equation}
that can be easily pointed out by using  Eq. (\ref{H3}). We must stress that this term is due to the dS expansion as we have shown for the free fermions on the dS expanding universe \cite{CdS2}.  The remaining real part 
\begin{eqnarray}
\Re E(\zeta)&=&\frac{\omega \zeta^2}{\pi}\left[K_{\nu_+}(i \zeta)K_{\nu_+}(-i\zeta)+K_{\nu_-}(i \zeta)K_{\nu_-}(-i\zeta) \right]\nonumber\\
&+&\frac{m \zeta}{\pi}\left[K_{\nu_+}(i \zeta)K_{\nu_-}(-i\zeta)-K_{\nu_-}(i \zeta)K_{\nu_+}(-i\zeta) \right]
\end{eqnarray}
is symmetric, $\Re E(-\zeta)=\Re E(\zeta)$, and satisfies the natural rest condition $\lim_{\zeta \to 0}\Re E(\zeta)=m$.

Furthermore, we focus on the partial currents starting with the general definition of the current density in a state $\psi$, 
\begin{equation}
I_{\psi}(t_c)=R(-\omega t_c)^{-3}\overline{\psi}(t_c,\phi)\gamma^{\phi}\psi(t_c,\phi)\,,
\end{equation}
and denoting 
\begin{equation}
I_{\lambda}^{\pm}(t_c)=R (-\omega t_c)^{-3}\overline{U}_{\lambda}^{\pm}(t_c,\phi)\gamma^{\phi} {U}_{\lambda}^{\pm}(t_c,\phi)\,.
\end{equation}
Hereby we obtain the {partial} currents,   
\begin{equation}
I_{\lambda}^+(t_c)= I^-_{\lambda}(t_c)=I_{\lambda}(t_c)\,,
\end{equation}
which are independent on polarization,  depending on time and $\lambda$ by means of $\zeta$ as $I_{\lambda}(t_c)=I[\zeta(t_c,\lambda)]$ or $I_{\lambda}(t)=I[\zeta(t,\lambda)]$ where
\begin{equation}\label{Il}
I(\zeta)=\frac{\zeta}{2\pi^2 R}\left[ K_{\nu_+}(i\zeta)K_{\nu_+}(-i\zeta)+
K_{\nu_-}(i\zeta)K_{\nu_-}(-i\zeta)\right]\,,
\end{equation}
is a real-valued skew-symmetric function, $I(\zeta)=-I(-\zeta)\in {\Bbb R}$.

These results can be applied to the state of arbitrary polarization $\psi_{\lambda}$ defined by Eq. (\ref{psil}) since  we have the following useful identities 
\begin{eqnarray}
&&\langle U_{\lambda}^{\pm},H_{AB}U_{\lambda}^{\mp}\rangle=0\,,\\
&&{\overline{U}_{\lambda}^{\pm}}(t_c,\phi)\gamma^{\phi}U_{\lambda}^{\mp}(t_c,\phi)=0\,.
\end{eqnarray}
which guarantee that  
\begin{eqnarray}
\langle \psi_{\lambda} H_{AB} \psi_{\lambda}\rangle& =& E_{\lambda}(t_c)\left(|c_+|^2+|c_-|^2\right)=E_{\lambda}(t_c)\,,\\
I_{\psi_{\lambda}}(t_c)&=&I_{\lambda}(t_c)\left(|c_+|^2+|c_-|^2\right)=I_{\lambda}(t_c)\,.
\end{eqnarray} 
In contrast to the flat case \cite{CAB1}, these quantities  are no longer conserved, evolving in time during the dS expansion. Nevertheless, we have the surprise to verify the fundamental identity of the AB effect, 
\begin{equation}\label{IE}
I(\zeta)=\frac{1}{2\pi R\omega}\frac{\partial \Re E(\zeta)}{\partial \zeta}~ \to~    I_{\lambda}(t_c)=-\frac{1}{2\pi \omega t_c}\frac{\partial E_{\lambda}(t_c)}{\partial \beta}\,,
\end{equation}
which is similar to that we deduced in the Minkowski spacetime \cite{CAB1} since in the flat limit we must take $\omega\to 0$ and $(-\omega t_c)\to 1$. The profiles of these functions are presented in Fig. 1 laying out the saturation of the partial currents for $\zeta\to \pm\infty$. We must stress that this is a genuine relativistic effect, as in the case of the AB rings in the Minkowski spacetime  \cite{CAB1}. The asymptotes coincide with the partial currents of massless fermions which are proportional with ${\rm sign}(\zeta)$.

{ \begin{figure}
    \centering
    \includegraphics[scale=0.50]{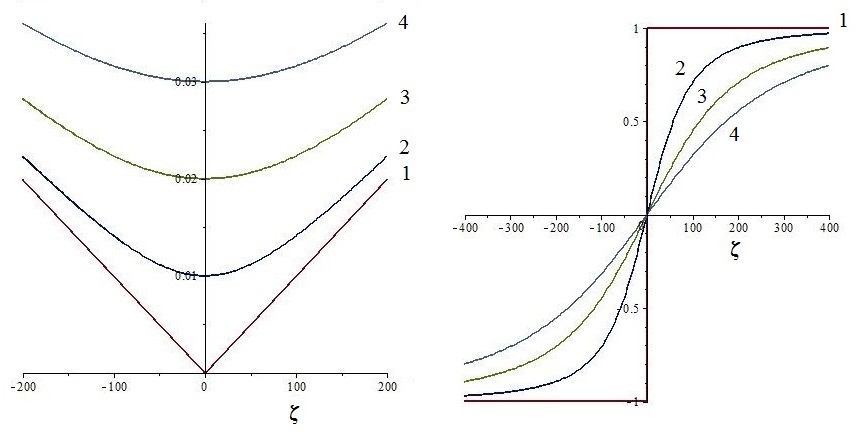}
    \caption{The functions $\Re E(\zeta)$ (left panel) and $2\pi R I(\zeta)$ (right panel) for $\omega=10^{-4}$ and different masses: $m_1=0$, $m_2=0.01$, $m_3=0.02$ and $m_4=0.03$.  }
  \end{figure}}

The last task is  to derive the total persistent current at $T=0$ in a semiconductor ring of radius $R$ having an even number of electrons, $N_e$, requested by the Fermi-Dirac statistics.  The total persistent current at $T=0$, written in terms of the proper time $t$ instead of the conformal one,  is given by the sum  
\begin{equation}\label{Isum}
{\cal I}(t)= \sum_{\lambda=-\lambda_F}^{\lambda_F}I_{\lambda}(t)= \sum_{\lambda=\frac{1}{2}}^{\lambda_F}\left(I_{\lambda}(t)+I_{-\lambda}(t)\right)
\end{equation}
over all the allowed angular numbers, $\lambda=\pm\frac{1}{2},\pm\frac{3}{2},...,\pm\lambda_F$ up to those of the pair of electrons of the Fermi level for which  $\lambda_F=\frac{1}{2}(N_e-1)$.  The flux parameter $\beta$ remains very small (less than $10^{-8}$) such that we can neglect the terms of the order $O(\beta^2)$ of the Taylor expansions of our functions that depend on  $\lambda+\beta$ by means of $\zeta$. Moreover, the first term of this expansion vanishes since 
\begin{equation}
\left[I_{\lambda}(t)+I_{-\lambda}(t)\right ]_{\beta=0}=\left[I(\zeta)+I(-\zeta)\right ]_{\beta=0}=0
\end{equation}
since $I(-\zeta)=-I(\zeta)$.  Thus we may approximate
\begin{equation}
I_{\lambda}(t)+I_{-\lambda}(t)\simeq 2 \beta \left.\frac{\partial I_{\lambda}(t)}{\partial \beta}\right|_{\beta=0}=2\beta \frac{\partial  I(\zeta_0)}{\partial \zeta_0}\frac{e^{-\omega t}}{\omega R}
\end{equation} 
where now $\zeta_0=\zeta|_{\beta=0}=\frac{\lambda}{\omega R}e^{-\omega t}$. As we proceeded  in the case of the AB rings in Minkowski spacetime \cite{CAB1},  we assume that the sum (\ref{Isum}) is well-approximated by the integral,
\begin{equation}
{\cal I}(t)\simeq2\beta \int_{\lambda=0}^{\lambda_F} \frac{\partial  I(\zeta_0)}{\partial \zeta_0}\frac{e^{-\omega t}}{\omega R}\,d\lambda=2\beta \int_{\lambda=0}^{\lambda_F}\frac{\partial  I(\zeta_0)}{\partial \zeta_0}\,d\zeta_0\,,
\end{equation}
arriving thus at the final expression of the persistent current at $T=0$,
\begin{equation}\label{Ifi}
{\cal I}(t)\simeq 2\beta  I[\zeta_F(t)]\,, \quad \zeta_F(t)=\frac{\lambda_F}{\omega R}e^{-\omega t}\,.
\end{equation}
where the function $I(\zeta)$ is given by Eq. (\ref{Il}). Moreover, if $\beta \ll \lambda_F$ we can write the corresponding energy of the Fermi level as
\begin{equation}\label{Efi}
E_F(t)\simeq\Re E[\zeta_F(t)]\,,
\end{equation}
neglecting the flux parameter $\beta$.

\section{Approximation}

The final exact results (\ref{Ifi}) and (\ref{Efi}) are expressed in terms of the complicated functions (\ref{El}) and (\ref{Il}) whose physical meaning is not quite obvious. For this reason we believe that it is convenient to use a convenient approximation  that can be easily understood and manipulated. An useful guide may be the asymptotic behavior of the energy and partial currents for  $\zeta\to \pm \infty$ that can be calculated according to the asymptotic form of the modified Bessel functions (\ref{Km0}) as,
\begin{equation}\label{asy}
I(\zeta)\sim \pm \frac{1}{2\pi R}\,, \quad \Re E(\zeta)\sim \omega| \zeta|\,. 
\end{equation} 
The limit of the function of the partial currents is independent $\zeta$ having just the asymptotic values we found in the case of the Minkowski spacetime \cite{CAB1}. 

These  results encourage us to replace the exact modified Bessel functions by the {\em pre-asymptotic} approximation (\ref{Bap}) obtaining the approximative formulas of our basic quantities,  
\begin{eqnarray}
\Re E(\zeta)& \sim& \tilde E(\zeta)=\sqrt{m^2+\omega^2\zeta^2}\,,\label{Ea}\\
I(\zeta)& \sim& \tilde I(\zeta)=\frac{1}{2\pi R}\frac{\omega\zeta}{\sqrt{m^2+\omega^2 \zeta^2}}\,,\label{Ia}
\end{eqnarray} 
which  preserve the identity (\ref{El}) since
\begin{equation}\label{IE}
\tilde I(\zeta)=\frac{1}{2\pi R\omega}\frac{\partial \tilde E(\zeta)}{\partial \zeta}~ \to~ \tilde I_{\lambda}(t_c)=-\frac{1}{2\pi  \omega  t_c}\frac{\partial \tilde E_{\lambda}(t_c)}{\partial \beta}\,.
\end{equation}
Note that this approximation is inspired by the {\em uniform expansion} of the modified Bessel functions which is proved for real or pure imaginary indices \cite{Or,NIST} but not for our complex indices $\nu_{\pm}$.  Therefore, we must verify the accuracy of this approximation resorting to numerical methods. For this purpose we consider the relative differences,
\begin{equation}
\delta E(\zeta,\kappa)=1-\frac{\tilde E(\zeta)}{\Re E(\zeta)}\,, \quad \delta I(\zeta,\kappa)=1-\frac{\tilde I(\zeta)}{I(\zeta)} \,,
\end{equation}  
which are functions on $\zeta$ and $\kappa=\frac{m}{\omega}$. Both these variables take large values since these are inverse proportional with $\omega$ which is extremely small even in strong gravitational fields (in our expanding universe  $\omega\simeq 2.5\, 10^{-18} s^{-1}$). By plotting  these relative differences on large domains of variables we obtain only very small  relative errors  (less than $10^{-6}$ as in Fig.2)  leading to the conclusion that Eqs. (\ref{Ea}) and (\ref{Ia}) represent a satisfactory {\em pre-asymptotic} approximation which is suitable for physical interpretation.

{ \begin{figure}
    \centering
    \includegraphics[scale=0.50]{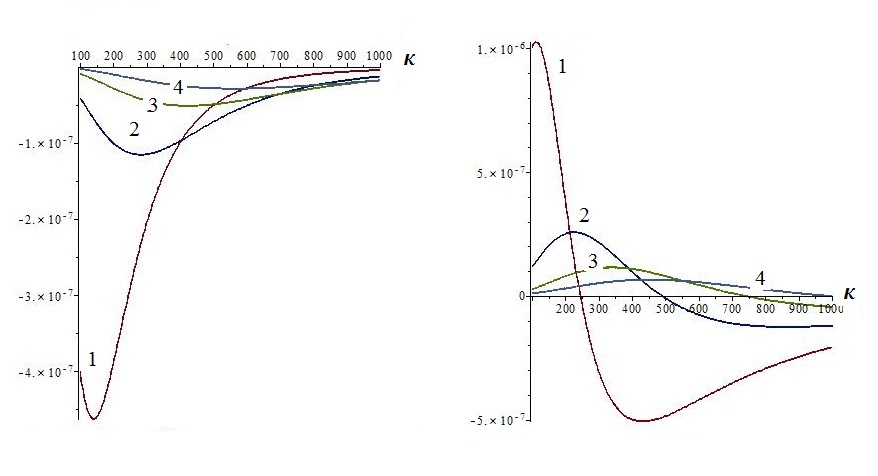}
    \caption{ The relative differences $\delta E$ (left panel) and  $\delta I$ (right panel) as functions of $\kappa=\frac{m}{\omega}$ for different values of $\zeta$:  $\zeta_1=200$,  $\zeta_2=400$,   $\zeta_3=600$,  and $\zeta_4=800$. }
  \end{figure}}

Trusting now in our approximation we may use it for deriving simpler and intuitive formulas by replacing the exact functions $I(\zeta)$ and $E(\zeta)$ by the functions (\ref{Ia}) and (\ref{Ea}). Then we obtain the final estimation
\begin{eqnarray}
\tilde{\cal I}(t)&\simeq & I_{max} \frac{\Lambda e^{-\omega t}}{\sqrt{1+\Lambda^2 e^{-2\omega t}}}\,,\quad I_{max}=\frac{\beta}{\pi R}\,,\label{Ifin}\\
\tilde E_F(t)&\simeq & m\sqrt{1+\Lambda^2 e^{-2\omega t}}\,,\label{Efin}
\end{eqnarray}
where the variable 
\begin{equation}
\Lambda= \frac{\lambda_F}{m R}\,,
\end{equation} 
is the same as in the flat case \cite{CAB1}  which can be recovered for $\omega \to 0$ since 
\begin{eqnarray}
\lim_{\omega \to 0}\tilde{\cal I}(t)&=& I_{max} \frac{\Lambda }{\sqrt{1+\Lambda^2 }}={\cal I}_{flat}\,,\\
\lim_{\omega \to 0}\tilde{E}_F(t)&=& m\sqrt{1+\Lambda^2}=E_{F\,flat}\,.
\end{eqnarray} 
Another advantage is of this simple time-dependence is that, around the initial moment $t=0$,   the small time-dependent correction                    
\begin{equation}\label{cor}
\tilde{\cal I}(t)={\cal I}_{flat}\left(1-\frac{\omega t}{1+\Lambda^2}\right)+{\cal O}(t^2)\,,
\end{equation}
due to the dS expansion can be pointed out. In other respects,  it is obvious that for $t\to \infty$ the AB effect vanishes since $\lim_{t\to \infty}\tilde {\cal I}(t)=0$ and $\lim_{t\to \infty}\tilde {E}_F(t)=m$.

\section{Concluding remarks}

We presented here the complete theory of the general relativistic AB effect in the dS expanding universe obtaining closed formulas of the persistent current and corresponding energy of the Fermi level. 

These quantities depend on the parameter $\Lambda$ that is independent on gravity being  determined by number of electrons $N_e$ (giving $\lambda_F=\frac{N_e-1}{2}$) and the 
parameter $m_e R$ (or $\frac{m_e c}{\hbar R}$ in usual units) where $m_e$ is the electron mass. For the mesoscopic systems $m_e R$ is very large ($\sim 10^4-10^6$) such that the parameter $\Lambda$ becomes very small remaining in the non-relativistic limit,
$I_{nr}\sim I_{max}\Lambda$, in which the saturation of the persistent current cannot be observed. This effect,  which is similar to that we found in special relativity  \cite{CAB1}, could be seen only for smaller values of $m_e R$ (say around $10^3$) that may be reached either in semiconductor mesoscopic rings where the effective electron mass $m^*$ is very small (e. g. $m^*<10^{-3}m_e$) or in semiconductor nano-rings. These conditions could be obtained in further experiments but now we cannot point out the relativistic effect of the saturation of the persistent current neither in the Minkowski spacetime nor in the dS one.

In what concerns the possibilities of measuring the time dependence of the persistent currents due to the expansion of our universe we do not have any hope since the time correction in Eq. (\ref{cor}) is proportional with $\omega$ which is too small for the actual experimental capabilities.

However, despite of the fact that the special or general relativistic AB effects cannot be measured actually in our laboratories, these effects do exist and, for this reason, these must be studied by using appropriate mathematical methods. In this manner, we may approach to a general relativistic solid state physics which could offer one new methods of investigating the local properties of the gravitational fields.       

\appendix

\section{Modified Bessel functions}

According to the general properties of the modified Bessel functions, $I_{\nu}(z)$ and $K_{\nu}(z)=K_{-\nu}(z)$ \cite{NIST}, we
deduce that those used here, $K_{\nu_{\pm}}(z)$, with
$\nu_{\pm}=\frac{1}{2}\pm i \mu$ are related among themselves through
\begin{equation}
H^{(1,2)}_{\nu}(z)=\mp\frac{2i}{\pi}e^{\mp \frac{i}{2}\pi\nu}K_{\nu}(\mp iz)\,, \quad z\in {\Bbb R}\,.
\end{equation} 
The functions  used here, $K_{\nu_{\pm}}(z)$ with
$\nu_{\pm}=\frac{1}{2}\pm i \mu$  ($ \mu \in {\Bbb R}$), are related among themselves through
\begin{equation}\label{H1}
[K_{\nu_{\pm}}(z)]^{*}
=K_{\nu_{\mp}}(z^*)\,,\quad \forall z \in{\Bbb C}\,,
\end{equation}
satisfy the equations
\begin{equation}\label{H2}
\left(\frac{d}{dz}+\frac{\nu_{\pm}}{z}\right)K_{\nu_{\pm}}(z)=-K_{\nu_{\mp}}(z)\,,
\end{equation}
and the identities
\begin{equation}\label{H3}
K_{\nu_{\pm}}(i z)K_{\nu_{\mp}}(-i z)+ K_{\nu_{\pm}}(-i z)K_{\nu_{\mp}}(i z)=\frac{\pi}{| z|}\,,\quad z\in{\Bbb R}\,.
\end{equation}
For $|z|\to \infty$ these functions behave as  \cite{NIST}
\begin{equation}\label{Km0}
I_{\nu}(z) \to \sqrt{\frac{\pi}{2z}}e^{z}\,, \quad K_{\nu}(z) \to K_{\frac{1}{2}}(z)=\sqrt{\frac{\pi}{2z}}e^{-z}\,,
\end{equation} 
regardless the index $\nu$.

The asymptotic approximation is rough since it looses the dependence on index of the functions $K_{\nu}$. For this reason we propose the pre-asymptotic approximation,
\begin{equation}\label{Bap}
K_{\frac{1}{2}\pm i\nu}(i\nu z)\sim\sqrt{\frac{\pi}{2\nu z}}\frac{\sqrt{\sqrt{1+z^2}\pm 1}}{(1+z^2)^{\frac{1}{4}}}\,e^{i\Theta(z)}\,,\quad z,\nu\in {\Bbb,R}\,,
\end{equation}
where $\Theta(z)$ remains un-defined. This approximation is inspired by the  uniform expansion of the modified Bessel functions which is proved only for real or pure imaginary indices \cite{Or,NIST}.

\end{document}